# Strategies for accessing the multi-pulse regime of mode-locked fiber lasers

P. Tchofo Dinda,[*] A. Malfondet, P. Grelu, and G. Millot[†]

*Laboratoire Interdisciplinaire Carnot de Bourgogne, UMR 6303 CNRS, Université de Bourgogne, 9 Av. A. Savary, B.P. 47870, 21078 Dijon Cedex, France*

A. Kamagate

*Department Mathématiques-Physique-Chimie, Université de Peleforo Gon Coulibaly, Côte d'Ivoire* (Dated: February 4, 2023)

Important ongoing research on mode-locked fiber lasers aims at developing new types of multi- soliton regimes, such as soliton molecules, molecular complexes or soliton crystals. The on-demand generation of such multi-pulse structures is a major challenge, whereas experiments generally in- volve a tedious trial-and-error adjustment of the laser parameters. Here, we present an approach based on a gradual and calibrated adjustment of the system configuration, which employs efficient parameter routes to reach well-defined multi-pulse regimes. Our numerical simulations show that once the mode-locking threshold is reached, we can adjust the laser parameters gradually to force the evolution of the laser dynamics towards multi-pulse structures with fewer distortion in their intensity profiles, which are accessible at reduced pump power levels.

## I. INTRODUCTION

Mode-locked fiber lasers (MLFLs) are highly valued for their ability to generate pulses whose intensity pro- files can be shaped by using tunable intracavity com- ponents, such as saturable absorbers, phase masks, dispersion compensators, or optical filters. However, the deterministic shaping of complex multiple-pulse profiles remains a major challenge. These complex waveforms notably include soliton molecules [1–7], macromolecules [8– 10], molecular complexes [11, 12], and soliton crystals [13, 14]. Soliton molecules, namely compact aggregates of few pulses bound by phase-sensitive interactions, are currently attracting a major attention for their potential applications in optical information encoding. Such applications require the generation of on-demand soliton molecules, which involves a complex control of intracavity parameters [15, 16]. However, to generate any of these complex multi-pulse structures, the first task is to efficiently access the multi-pulse regime of the laser. The pumping power (PP) is the most accessible and determining cavity parameter in that process. The mode-locking threshold is defined as the minimum PP needed to access any short-pulse regime. When this threshold is reached, we assume that the laser generates a single pulse inside the cavity. This condition is not always met, in particular for fiber laser setups designed for a low pulse energy and high-harmonic mode locking operation [17, 18]. Nevertheless, we can assume that this is generally the case for a fiber laser designed to generate just a few pulses. After the single-pulse regime is reached, by increasing the PP further, a multi-pulsing instability develops so that the laser switches to the multi-pulse regime [19–24]. In the following, we will refer to this common procedure of accessing the multi-pulse regime as the conventional approach.

In some laser configurations, the PP needed to generate multi-pulse structures can be much higher than the mode locking threshold so that the pulses can be subjected to significant distortions in their intensity profile [25]. The framework of dissipative solitons explains how the field profile of any stable pulse structure is strongly affected by a combination of dispersive and dissipative, linear and nonlinear, propagation effects [9]. Spectral filtering is one of the most important effects, which affects the overall laser dynamics and the stationary pulse profile. Whereas the finite bandwidth of the gain profile of the doped fiber has already an important effect on the pulsed laser dynamics [26, 27], its impact can be strongly enhanced if a bandpass filter (BPF) with a bandwidth lower or of the same order as that of the gain bandwidth, is inserted into the cavity [28–31]. Therefore, many previous studies have used BPFs to modify the laser dy- namics. For instance, it has been shown that BPFs can be used to design MLFLs with large tunability capabilities, such as the ability to adjust the pulse wavelength [32], adjust separately the wavelength and the temporal width of pulses [33–35], stretch and compress the pulses in the temporal domain [36, 37], or compress the pulses in the spectral domain to generate their fragmentation [23, 24, 38, 39]. Another class of applications of filtering, is to sculpt the intensity profile of stable states, such as parabolic-, flat-top-, triangular- and saw- tooth-profiled pulses [40–42], dispersion-managed soliton profiles [43], similaritons [39], and rogue waves [44]. At this point, we need to raise the issue of hysteresis and multistability, where in regions of the parameter space, the laser can generate several distinct stable states corresponding to the same cavity parameters [20]. The access to one of the stable states at a given time depends on the dynamical trajectory followed by the laser when the cavity pa-rameters are varied to reach their target values. In most experiments, such trajectories are not known in advance, which is why a trial and error procedure generally precedes the generation of complex multi-pulse structures.

--------

[*] tchofo@u-bourgogne.fr
[†] Also at Institut Universitaire de France, 1 Rue Descartes 75005 Paris, France

In the present study, we consider a MLFL with adjustable pulse generation capacities provided by two tunable laser components: an erbium-doped fiber amplifier (EDFA) and a BPF with tunable bandwidth. We show that by following a specific procedure to adjust the PP of the EDFA and the bandwidth of the BPF, the laser dynamics follows a trajectory leading to stable multi-pulse states with pre-defined properties. In comparison with situations where the conventional approach causes huge distortions in the pulse's intensity profile, we demonstrate here that our procedure allows to reach multi-pulse regime and generate pulses without profile distortions, at reduced pump power. Our approach is simple to implement as it combines tunable components commercially available.

## II.   NUMERICAL MODELING

Our laser model is a unidirectional ring cavity that combines the following four major components placed in the order indicated in Fig. 1: a bandpass filter (BPF) with tunable bandwidth, a section of Dispersion-Compensating Fiber (DCF) to adjust the average cav- ity dispersion, a section of Erbium Doped fiber (EDF, pumped by a laser diode through a WDM coupler), and a SESAM (SEmi-conductor Saturable Absorber Mirror) used to insert nonlinear losses and favor laser mode locking. Two output couplers (OC), labelled OC1 and OC2 respectively, are inserted into the cavity. The coupler OC1 is placed after the output of the EDFA, where the pulse energy is maximum for the user. The coupler OC2, placed at the input of the BPF, is used to characterize the pulse before its passage through the BPF. A circulator (CIR) is placed between the couplers OC1 and OC2 to receive the light reflected by the SESAM and inject it back into the cavity. The circulator also imposes a unidirectional propagation of light through the cavity. Since all components are spliced to single-mode fibers (SMF), the cavity is thus equipped with an active fiber and a number of sections of SMF. In our scalar model, the pulse propagation in these fiber sections can be described by the following generalized nonlinear Schrodinger equation (GNLSE)[23, 45, 46]:

$$\psi_z + \frac{i\beta_2}{2}\psi_{tt} - \frac{\beta_3}{6}\psi_{ttt} + \frac{\alpha - g(z, P_{av}, \nu_s)}{2}\psi = i(1-\rho)\gamma|\psi|^2\psi + R[\psi] \quad (1)$$

where $\beta_2$, $\beta_3$, $\gamma$, $g$, and $\alpha$ designate the second-order dispersion (SOD),third-order dispersion (TOD), Kerr non-linearity, gain, and linear attenuation parameters, respectively. The term $R[\psi]$, which describes the Raman effect, is written as follows [46]: $R[\psi] = i\gamma\rho\psi\int_0^\infty \chi_R(s)|\psi|^2(t-s)ds$, where $\chi_R$ represents the Raman susceptibility, and $\rho$ is the fractional contribution of the Raman scattering to the total nonlinearity, with $\rho = 0.18$ [46]. In Eq. (1), $g = 0$ in the case of the passive fiber sections (SMF), whereas for the active fiber (EDF) $g(z, P_{av}, \nu_s)$ is calcu-lated at any signal position $\nu_s$, from the rate equations for the pump and signal power at a given longitudinal position z along the active fiber [23]. Here, $P_{av}$ is the average power at the input facet of the EDF.
The GNLSE (1) is numerically solved by means of the split-step Fourier method [45].

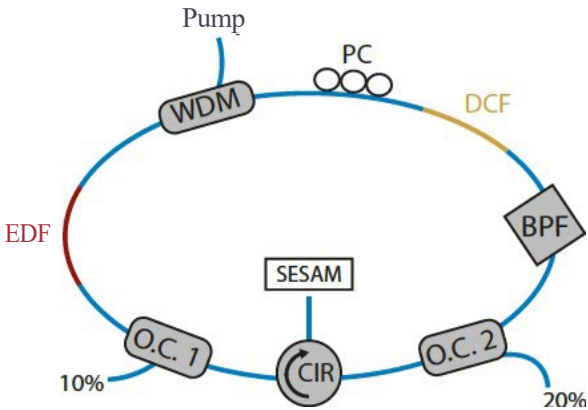

FIG. 1. Schematic of the fiber laser cavity.

The action of the SESAM on an incident light field is modeled as instantaneous. This would correspond in practice to a SESAM with a sub-picosecond response time applied on picosecond laser pulses. It is modeled by the following monotonous transfer function for the optical power:

$$P_o = T P_i, \quad (2a)$$
$$T \equiv T_0 + \Delta T P_i/(P_i + P_{sat}), \quad (2b)$$

where $T$ describes the transmission of the instantaneous SA, $T_o$ is its transmitivity at low signal, and $b.T$ the absorption contrast, while $Pi$ $(P_o)$ designates the instantaneous input (output) optical power.

The BPF spectral profile is modeled by the following super-Gaussian function:

$F_{BPF}(\omega) = \exp\left[-2^{2m}\log(2)\frac{\omega^{2m}}{\Delta\Omega_{BPF}^{2m}}\right]$,   where is the filter's bandwidth, while $m$ is an integer that determines the filter's profile. The value $m = 1$ corresponds to a Gaussian profile, whereas for $m > 1$ the BPF's profile has an increasingly flat top and steeper flanks as m increases. Throughout the present work, we model the flat-top filter with $m = 4$. The other lumped elements of the laser cavity (coupler, splices) only induce broadband linear losses. Our numerical simulations are performed using the following parameters:

- **EDF:** SOD : -15 ps nm$^{-1}$ km$^{-1}$, TOD : $\sim$ 0 ps nm$^{-2}$ km$^{-1}$, Effective mode area = 28.3 $\mu m^2$, Length = 1.2 m, Loss = 0.2 dB/km.

- **SMF:** SOD : 18 ps nm$^{-1}$ km$^{-1}$, TOD : 0.07 ps nm$^{-2}$ km$^{-1}$, Effective mode area = 78.5 $\mu m^2$, Total length = 15.81 m, Loss = 0.2 dB/km.

- **DCF:** SOD : -91.7 ps nm$^{-1}$ km$^{-1}$, TOD : -0.12 ps nm$^{-2}$ km$^{-1}$, Effective mode area : 20 $\mu m^2$, Length: 2.9 m, Loss : 0.6 dB/km.

- **Saturable absorber:** $T_0 = 0.7$ (70%), $\Delta T = 0.3$ (30%), $P_{sat} = 10$ W.

- **Output coupler transmissions:** $OC_1$ (90%), $OC_2$ (80%) .

## III. CONVENTIONAL PROCEDURE OF ACCESS TO THE MULTI-PULSE REGIME

The conventional approach to access the multi-pulse regime of a MLFL is to first adjust the laser PP slightly above the mode-locking threshold to generate a single pulse in the cavity. Then, the PP is increased gradually until the single pulse present in the cavity becomes un- stable and fragments into two (or more) pulses. Figure 2 illustrates, in color scale, the evolution of the temporal intensity of the intra-cavity electric field recorded at the input of the BPF, when the PP is increased from 21mW to 33mW. The panel (a1) of Fig. 2 shows that in a cavity equipped with a BPF of 4nm bandwidth, the switch to the multi-pulse regime occurs at a PP of 25.6mW, which leads to the fragmentation of the initial pulse into two pulses. By continuing to increase the PP, a new fragmentation process occurs at 29.3mW, followed by a re- structuring of the intracavity field into three pulses. The behavior observed for the cavity equipped with a BPF of 4nm bandwidth (panel 2 (a1)), differs drastically from that shown in panel 2 (a2) corresponding to a bandwidth of 12nm. Indeed, in the same PP range, i.e. between 21mW and 33mW, the multi-pulse regime is inaccessible with a 12nm-bandwidth filter. Comparison between the results of panels 2 (a1) and 2 (b1) shows that for

$\Delta\lambda = 4$nm, the flat-top filter increases the PP transition to the multi-pulse regime and reduces the number of frag- mentation points. For large filter bandwidths, there is no significant difference related to the type of filter spectral profile considered [see panels 2 (a2) and 2 (b2)].

Thus, the most striking feature of the conventional procedure for accessing the multi-pulse regime is the pulse fragmentation shown in panels (a1) and (b1) of Fig. 2. However, Fig. 2 gives no insight into the physical mechanism underlying the fragmentation process. To get an idea of this physical mechanism, we show in Fig. 3 the evolution of the main physical parameters of the pulse in the case considered in panels 2 (a1) and 2 (b1), where

$\Delta\lambda = 4$nm. Here and thereafter, the pulse parameters are calculated using collective coordinate techniques

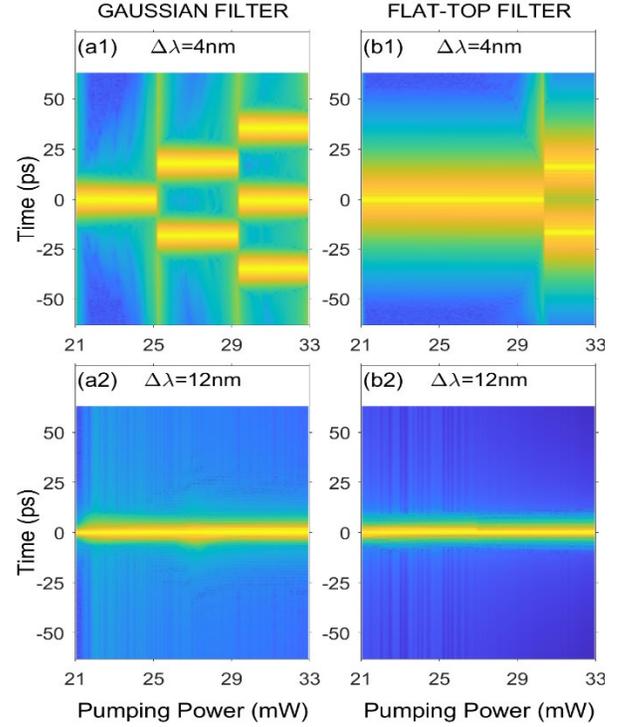

FIG. 2. Evolution in color scale of the temporal intracavity intensity as a function of the pumping power. (a1)-(b1) : $\Delta\lambda = 4$nm. (a2)-(b2) : $\Delta\lambda = 12$nm.

[22, 46]. Panels (b1) and (b2) of Fig. 3 show that the total intra-cavity energy (including all pulses) increases monotonically with PP and makes a steep jump at each fragmentation point. Panels 3 (c1) and 3 (c2) show that the pulse energy increases monotonically with PP and falls sharply at each fragmentation point. The pulse's peak power evolves in the same way as the energy, as illustrated by panels 3 (d1) and 3 (d2). Unlike the evolution of peak power, the temporal width of the pulse decreases monotonically as the PP increases and makes a sharp jump at each fragmentation point. This temporal narrowing of the pulse in the phase preceding the fragmentation point is in line with the widening the pulse spectrum, as shown by the panels 3 (f1) and 3 (f2). This turns out to be the essential pulse splitting mechanism: as we gradually increase the PP, the pulse spectrum widens up to a critical point close to the filter bandwidth, where the action of the filter destabilizes and fragments the pulse. It should be noted that each fragmentation phenomenon causes a steep variation of the value of all pulse parameters. This steep variation can be easily understood if we make an analogy between the pulse behavior and that of an object that we try to pass through a window. If the size of the object is smaller than that of the window, the object will pass through the window without altering its physical integrity. On the other hand, if the size of the object is larger than that of the window, the object will pass through the window only if it is frag-

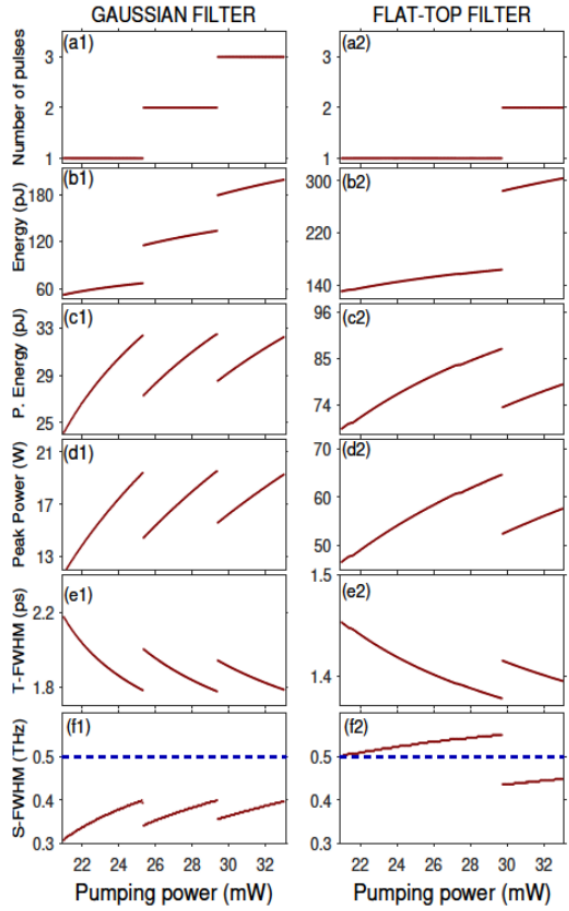

FIG. 3. Evolution of the pulse parameters as a function of the pumping power, for $\Delta\lambda = 4$nm. The pulse parameters are calculated from the pulse profile at the BPF input. (a1)-(a2) : Number of pulses generated. (b1)-(b2) : Total energy of the intra-cavity field. (c1)-(c2) : Pulse energy. (d1)-(d2) : Pulse peak power. (e1)-(e2) : Temporal FWHM (Full Width at Ralf Maximum of the peak intensity). (f1)-(f2) : Spectral FWHM.

mented into several pieces smaller than the window. This fragmentation phenomenon is most clearly visible in panels 3 (f1) and 3 (f2), where a sharp reduction in the pulse spectral width is observed. However, the specificity of the optical pulse is that it is subject to a condition of stability in the laser cavity, requiring a specific intensity profile to ensure the balance between the different physical phenomena present in the cavity (gain, loss, dispersion, non-linearity, etc.). Therefore, at the point of fragmentation, the reduction of the spectral width of the pulse is necessary but not sufficient. A restructuring of the entire pulse profile is required to satisfy the stability condition in the cavity, hence the abrupt change in all the physical parameters of the pulse, which is comparable to a relaxation process leading to pulse re-confinement in the filter bandwidth.

On the other hand, the panels (f1) and (f2) in Fig. 3 highlight a major difference between the two types of filters in their way of shaping the pulse intensity profile. Indeed, we see in panel 3 (f1) that the Gaussian filter acts in such a way as to confine the pulse spectrum strictly within its FWHM bandwidth (indicated by the horizontal line in dashes). Quite unexpectedly, despite its steep flanks, the fiat-top filter lea.ds to pulses whose spectra clearly exceeds the FWHM filter bandwidth, as shown in panel 3 (f2). This bandwidth overflow reaches its maximum just before the fragmentation point, i.e. at 29.7mW. This phenomenon, which was also reported in Ref. [27], is even clearer in Figure 4, in which the curves in solid lines represent the intensity profiles recorded at the filter input just before the first fragmentation point, i.e. at at 25.3mW (29.7mW) for the Gaussian (fiat-top) filter. In the panels 4 (b1) and 4 (b2), the dash curves represent the spectral profile of the filter, with a bandwidth of 4nm (equivalent to 0.5THz). The dotted curves in figure 4 illustrate the fact that the pulse spectrum does not exceed the limits of the filter bandwidth when the PP is sufficiently far from the first fragmentation point. As the PP rises and approaches the frag-mentation point, the pulse spectrum widens but with- out crossing the bandwidth limits of a Gaussian filter, while there is an overflow of the bandwidth in the cavity equipped with the fiat-top filter. In addition, careful examination of panels 4 (a1) and 4 (b1) reveals that the Gaussian filter favors the generation of pulses whose intensity profile has a bell curve shape, while the fiat-top filter inflicts slight profile distortions, which are more vis-ible in the panel 4 (b2) than in the panel 4 (a2).

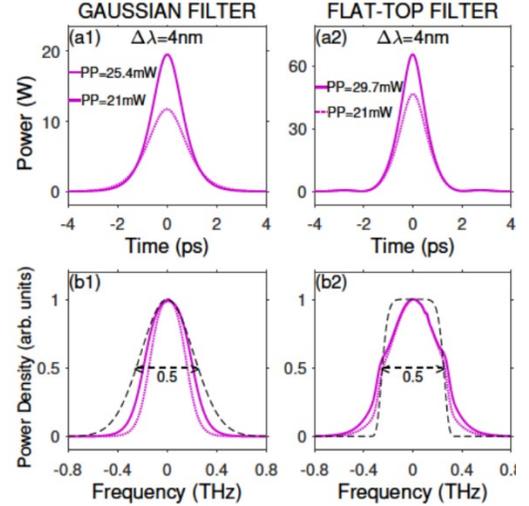

FIG. 4. Pulse intensity profile just before the first fragmentation point for the cavity considered in Fig.2. The dotted curve shows for comparison the pulse profile for a pumping power well before this fragmentation point. (a1)-(a2) : Temporal profile. (b1)-(b2) : Spectral profile.

In Fig. 3, we have highlighted the evolution of the pulse parameters as a function of PP, but with parameters evaluated at a specific point of the cavity, i.e., at the BPF input. As we consider a dispersion-managed cavity, i.e., a cavity where we have inserted a DCF fiber in order to obtain an average cavity dispersion value close to zero, the pulse parameters are likely to vary substantially over a cavity round-trip. The knowledge of this pulse's internal dynamics is useful to appreciate the respective roles of the different components of the cavity in the structuring of the stable states of the laser. The solid line curves in Fig. 5 illustrate the intra-cavity dynamics of the pulse parameters for a PP just before the fragmentation point, while the dotted curves illustrate the dynamics just after the fragmentation point. The dashed curves show the internal dynamics for a PP located well before the fragmentation point. Note that the curves of evolution of the energy and the peak power contain a number of small jumps corresponding to the linear losses between the different components of the laser. More importantly, Fig. 5 highlights the following major points:

(i) The PP located just before the fragmentation point (i.e., 25.4mW and 29.7mW respectively for the Gaussian

and flat-top filter) is the operating condition of the cavity for which the internal dynamics are strongest, as remarkably illustrated by panels (b2), (c2), (d2) and (e2) of Fig.

5, where we see that at each pass, the flat-top filter dramatically changes the pulse intensity profile, causing a drop of 51% in its peak power, a jump of 72% in its temporal FWHM, a drop of 24% in its spectral FWHM, and a drop of 61% in its spectral full width (FW). Here and thereafter, the FW is defined as the width at 1% of the maximum. This operating condition is also the one that exacerbates the bandwidth overflow phenomenon, because here, the overflow relates to the pulse's spectral FWHM. Indeed, the panel 5 (d2) shows that over a distance from z=2.25m to the entrance of the flat-top filter located at z=9.63m (which corresponds to almost one third of the length of the cavity), the spectral width of the pulse is clearly beyond the filter bandwidth.

(ii) The dotted curves in Fig. 5 show that a relaxation of the internal dynamics occurs just after the fragmentation point, with a major effect visible in the panel 5 (d2), which is to bring the spectral dynamics of the pulse back within the bandwidth of the system. On the other hand, relaxation does not suppress the spectral overflow as far as the spectral FW (full width) is concerned, as can be seen from panel 5 (e2), which shows that there is overflow of bandwidth over almost the entire length of the cavity, except for a distance of about 2 meters after crossing the flat-top filter.

As we have seen in Fig. 2, the PP of access to the multi-pulse regime via the conventional approach, defined as the first point of fragmentation, differs quantitatively depending on whether $\Delta\lambda$ =4nm or $\Delta\lambda$ =12nm. To have a more in-depth idea of the influence of filter bandwidth, we gradually raised the PP from the mode-

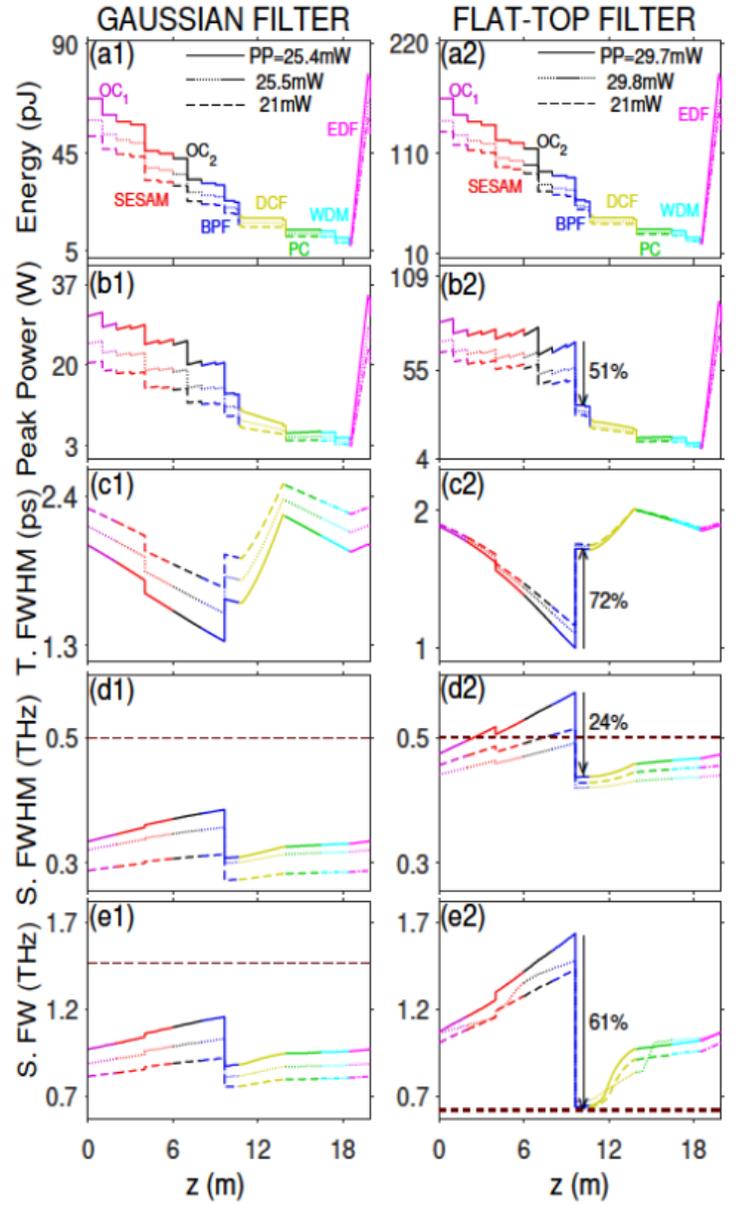

: Pulse peak power. (c1)-(c2) : Temporal FWHM. (d1)-(d2) : Spectral FWHM. (e1)-(e2) : Spectral FW (Full Width).

FIG. 5. Evolution of the pulse parameters over a roundtrip in the cavity for 5 nm. (a1)-(a2)

locking threshold to the first fragmentation point, for different values of $\Delta\lambda$ between 4nm and 12nm. The result is visible in the Fig. 6, where panel (a) shows the value of PP as a function of filter bandwidth, while the other panels represent the pulse parameters just before the fragmentation point. Panels 6 (a1) and 6 (a2) show that the evolution curve of the PP roughly resembles a two-level staircase. The resulting intra-cavity energy also has an evolution curve that resembles a staircase, as shown in panels 6 (b1) and 6 (b2). In panels 6 (a1) and 6 (a2), we observe that the PP varies only very slightly over a wide range of values of $\Delta\lambda$ ranging from 4nm to some critical value, say . $\Delta\lambda_c$, which is about 11nm. As soon as $\Delta\lambda$ exceeds $\Delta\lambda_c$, the PP abruptly jumps to a level that is almost double the PP level in the region

$\Delta\lambda < \Delta\lambda_c$. However, we observe that in the region where $\Delta\lambda > \Delta\lambda_c$, the PP fluctuates much more than

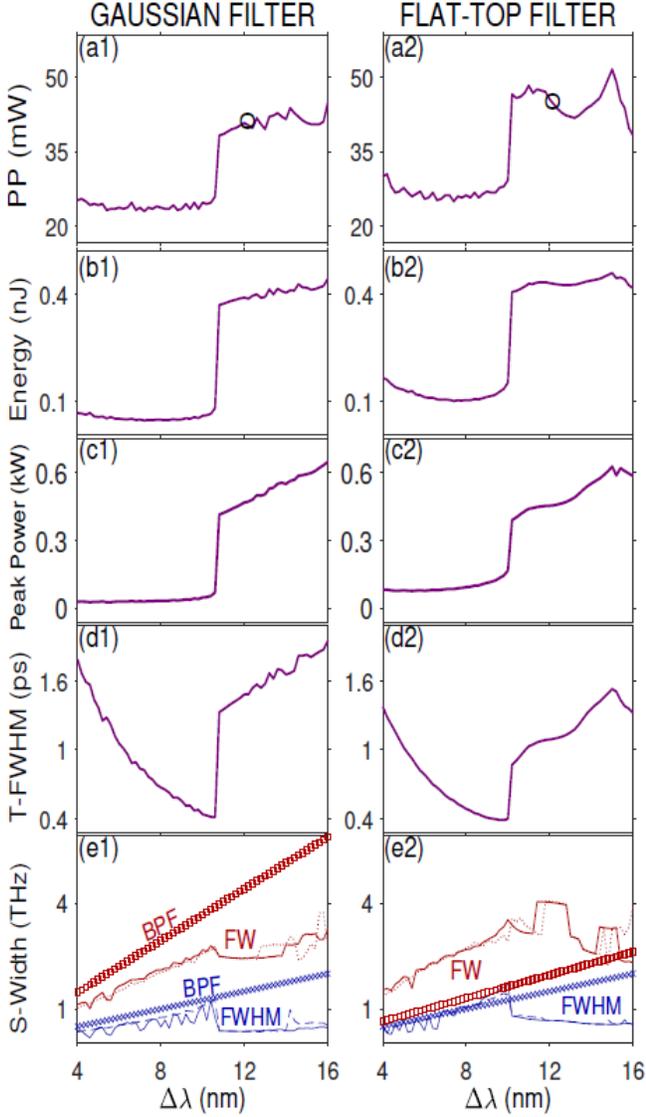

FIG. 6. Access to the multi-pulse regime through the conventional approach. (a1)-(a2) : PP of access to the multi-pulse regime as a function of filter bandwidth. (b1)-(b2) : Total energy of the intracavity field. (c1)-(c2) : Peak power. (d1)-(d2) : Temporal width. (e1)-(e2) : Spectral width.

in the region $\Delta\lambda < \Delta\lambda_c$, These large fluctuations in PP reflect less stability of the pulse, which can result from the combination of several factors. The first is the dramatic increase in non-linearity induced by the high level of the pulse's peak power, which is an order of magnitude larger than that we observe for $\Delta\lambda < \Delta\lambda_c$, as shown in panels 6 (c1) and 6 (c2). The second factor is probably related to the filtering effect induced by the gain curve of the active fiber, which is expected to come into play when the filter bandwidth is sufficiently large, i.e., in the region $\Delta\lambda > \Delta\lambda_c$, Unlike our BPF whose bandwidth is manually adjustable in a way completely independent of the operating conditions of the laser, the filtering induced by the gain curve of the active fiber is very sensitive to the average power of the optical field at the input of the active fiber, which is itself highly dependent on the pumping power of the active fiber. More importantly, one can raise the question of the physical origin of the critical behavior occurring at $\Delta\lambda = \Delta\lambda_c$, The answer to this question can be obtained by carefully examining the evolution curves of the pulse spectral widths in panels 6 (e1) and 6 (e2). In these panels, the lines represented by small cross and diamond symbols represent respectively the filter bandwidth at FWHM and at FW. Indeed, we observe for $\Delta\lambda < \Delta\lambda_c$, that the spectral width of the generated pulses follows almost the same ascending curve as that of the filter, thus indicating that the filter plays a crucial role in shaping the pulse profile. Beyond $\Delta\lambda_c$, the pulse spectral width stops growing, thus indicating that the BPF no longer plays a significant role in intra-cavity dynamics, and consequently, the filtering effect induced by the gain curve of the active fiber comes into play. But in our cavity, there is no way to control this filtering effect. The main features of the conventional procedure for accessing the multi-pulse regime of a MLFL are remarkably illustrated in Fig. 7, which shows the intensity profiles of the stable states obtained respectively for $\Delta\lambda = 8$ nm and $\Delta\lambda = 12$ nm, which correspond to two bandwidth values located respectively below and above the critical value $\Delta\lambda_c \sim 11$ nm revealed in Fig. 6. It should be remembered that for each of the two values of $\Delta\lambda$ considered, the procedure consists in gradually increasing the PP to a value slightly above the fragmentation point. The panels (a1)-(b1) and (a2)-(b2) in Fig. 7 clearly illustrate that for $\Delta\lambda = 8$ nm $< \Delta\lambda_c$, the multi-pulse state consists of pulses whose intensity profile has a nice bell-shaped curve, without significant distortions. On the contrary, for $\Delta\lambda > \Delta\lambda_c$, the pulse profiles are affected by very severe distortions, as is clearly shown by the inserts inside the panels (a3) and (a4) in Fig. 7. We attribute those profile distortions to the nonlinear effects induced by the much larger pulse's peak power in the region

$\Delta\lambda > \Delta\lambda_c$ (which is nearly twice as much as that of the pulses generated when $\Delta\lambda < \Delta\lambda_c$). Those pro- file distortions result from the high level of PP required to reach the multi-pulse regime, which is almost double the PP required when $\Delta\lambda < \Delta\lambda_c$, In the following section, we present a procedure for accessing the multi-pulse regime, which avoids the two disadvantages raising from the conventional approach.

## IV. ACCESSING THE MULTI-PULSE REGIME VIA A TWO-DIMENSIONAL PARAMETER APPROACH

In the previous section, we analyzed the conventional procedure for accessing the multi-pulse regime of MLFLs through the use of the PP as the only control parameter beyond the mode locking threshold. On the other hand, we have highlighted in Fig. 6, the major influence of the BPF's bandwidth on the PP that is needed to access the multi-pulse regime. The peculiarity of our 2D (two-dimensional) procedure for accessing the multi-pulse regime is to use the BPF's bandwidth $\Delta\lambda$ as a system configuration parameter, in the same way as the PP.

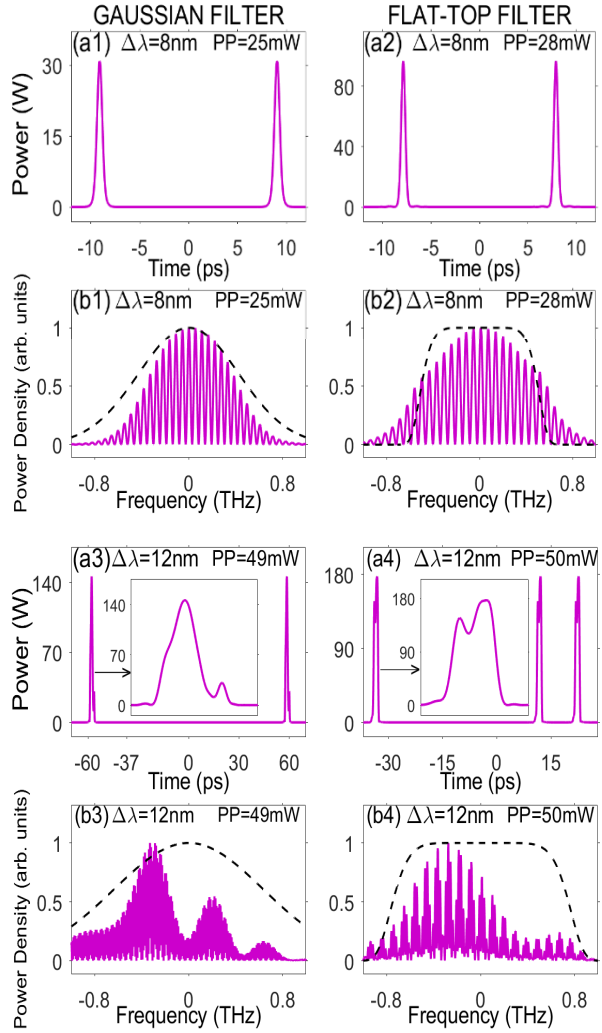

FIG. 7. Multi-pulse intensity profile obtained by the conventional approach, for a PP slightly above the first fragmentation point. (a1)-(b1) : Multi-pulse profile for $\Delta\lambda$ = 8nm for the cavity with Gaussian filter. (a2)-(b2) : Multi-pulse profile for $\Delta\lambda$ = 8nm for the cavity with flat-top filter. (a3)-(b3) : Multi-pulse profile for $\Delta\lambda$ = 12nm for the cavity with Gaussian filter. (a4)-(b4) : Multi-pulse profile for $\Delta\lambda$ = 12nm for the cavity with flat-top filter.

In this procedure, we shall move the PP and ad- just the filter's bandwidth, so as to minimize the PP needed to reach the multi-pulse regime. As we can see in panel (a1) of Fig. 6, in the conventional approach, the PP for accessing the multi-pulse regime is minimal for a filter bandwidth below $\Delta\lambda_c$, while for $\Delta\lambda > \Delta\lambda_c$ the PP is about twice higher. Therefore, the 2D approach should be particularly useful for easy access to the multi- pulse regime in cavities where $\Delta\lambda > \Delta\lambda_c$. To illustrate this procedure, consider again the laser cavity configured with a bandwidth of 12nm, while keeping in mind that for $\Delta\lambda$ =12nm the PP required to access the multi-pulse regime is greater than 40mW when the conventional approach is used, as indicated by the small circle-shaped symbols in panels (a1) and (a2) of Fig. 6. Our 2D approach takes place in two steps:

(i) First, we set the filter bandwidth to a value $\Delta\lambda_{opt}{=}\Delta\lambda$ for which the multi-pulse regime is easily accessible by the conventional approach, i.e., accessible with the lowest possible PP. Then, from the mode-locking thresh- old, we gradually increase the PP up to the value $P_{PF}$ where fragmentation occurs. In the second step of the two-dimensional procedure, the PP is maintained at the value $P_{PF}$ .

(ii) The second step begins with a laser already in multi-pulse regime but with a bandwidth $\Delta\lambda_{opt}$ = $\Delta\lambda$. Here, the procedure consists of gradually increasing the filter's bandwidth from $\Delta\lambda_{opt}$ to $\Delta\lambda$. We emphasize that such bandwidth adjustment must be done gradually in order to prevent the laser from exiting the multi-pulse regime.

Panels (a1) and (a2) of Fig. 8 show the results obtained using the 2D approach described above, starting the first step of the procedure with $\Delta\lambda_{opt}$ =8nm and ending just after the fragmentation point, at a PP of 24.50mW (27.35mW) for the cavity equipped with a Gaussian (flat-top) filter. Note that in Fig. 8 and all subsequent figures, the intra-cavity field is recorded at the BPF input. Thus, the panels 8 (a1) and 8 (a2) show that the first step of our 2D approach results in the generation of two pulses in the cavity, in a similar way to the conventional approach. In the second step of the procedure, we gradually increase the bandwidth from 8nm to 12nm. But in panel 8 (a1), during this operation, the two pulses remain stable up to a certain value of $\Delta\lambda$ close to 12nm, where a *defragmentation* phenomenon occurs and restructures the intra-cavity field so as to produce a single pulse in the cavity. Note that defragmentation is not clearly visible in Fig. 8 (a1) because it occurs at the very end of the wavelength range considered, i.e., 11.97nm. Thus, in the panel 8 (a1), our goal of accessing the multi- pulse regime in a cavity with a bandwidth of 12nm was not achieved by finally setting the PP to the level just after the fragmentation point in the first step of the procedure. In this context, the question arises as to whether we could access the multi-pulse regime by completing the first phase at a PP slightly greater than the fragmentation point. The answer to this question is given in the panel 8 (b1) where it is observed that, by ending the first step with a PP slightly above the fragmentation point, i.e. 25mW, we finally access the multi-pulse regime with a filter of 12nm of bandwidth, but here, three pulses are generated within the cavity.

In contrast, panel 8 (c1) shows that by ending the first step with a PP slightly below the fragmentation point, the laser enters the second phase of the 2D approach when it is still in single-pulse regime, but does not re- main in this regime throughout the rest of this phase. Against all odds, despite the PP being below the frag-mentation point, a fragmentation process occurs at a value of $\Delta\lambda$ ~9nm. The laser then enters a multi-pulse regime and remains there until $\Delta\lambda$ = 11.9nm, and there, a defragmentation occurs (but without being clearly visible in the panel 8 (c1), much like in the case of the panel 8 (a1)) and brings the laser back into single-pulse mode. Furthermore, we see in panels (a2), (b2) and (c2) of Fig. 8 that a cavity equipped with a flat-top filter shows a qualitative difference compared to a cav- ity equipped with a Gaussian filter. Indeed, in the three cases considered, i.e., a PP located just after the fragmentation point (panel 8 (a2)), slightly above the fragmentation point (panel 8 (b2)), and slightly below the fragmentation point (panel 8 (c2)), the laser enters a second multi-pulse regime that is non-stationary and comprises two pulses of unequal intensity. Such phenomenon is not observed with the Gaussian filter, which generates stationary pulses of the same intensity (panel 8 (b1)). Fig. 9 shows the intensity profile of the structure generated via the 2D procedure for accessing the multi-pulse regime described in Fig. 8, highlighting a difference in the number of pulses generated, i.e., three pulses in the case of the Gaussian filter and two pulses for the flat-top filter.

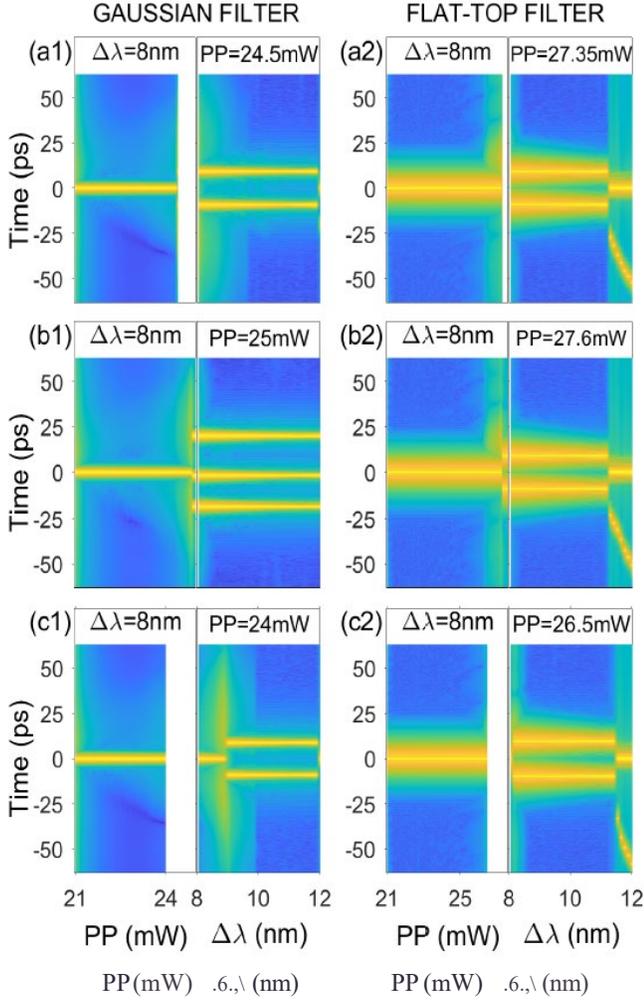

FIG. 8. Evolution of the temporal optical intensity along with the pumping power (PP) increase at $\Delta\lambda$ =8nm (first step), and with the filter bandwidth increase at fixed PP (second step). (a1)-(a2) : The final PP is set just after the fragmentation point. (b1)-(b2) : The final PP is set slightly above the fragmentation point. (c1)-(c2) : The final PP is set slightly below the fragmentation point.

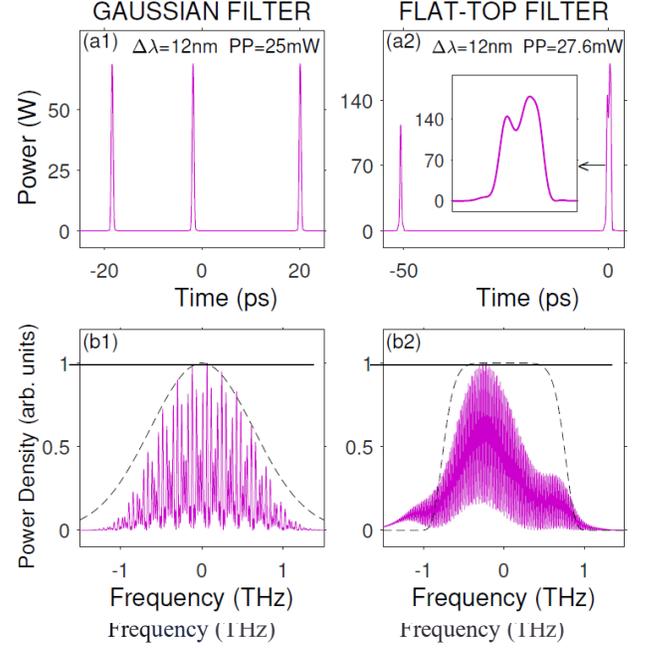

FIG. 9. Multi-pulse intensity profile generated at the end of the 2D procedure shown in panels (c1) and (c2) of Fig. 8. (a1)-(a2) : Temporal profile. (b1)-(b2) : Spectral profile.

We discovered in Fig. 8 that an undesirable phenomenon can occur during the configuration of the cavity by the 2D approach, namely, the defragmentation appears of the multi-pulse structure. Defragmentation appears as a critical phenomenon of restructuring of an intra-cavity field initially having a profile very far from that is really adapted to the configuration parameters of the cavity. In general, a restructuring phenomenon resulting in a change in the number of pulses within a laser cavity (fragmentation or defragmentation) is a very brief process during which the intra-cavity field executes fluctuations of large amplitude, temporarily requiring a large amount of energy. A gradual increase in PP naturally favors this type of process by an energy supply activated from outside the cavity. Now, in the second step of the 2D approach, we gradually increase the bandwidth of the BPF, which has the consequence of reconfiguring the cavity towards the single-pulse regime. But by keeping PP at a fixed level, we also deprive the cavity of an energy source capable of promoting a huge restructuring of the intra-cavity field. This is the reason why the system remains in multi-pulse mode over a wide range of values of $t..>.$, before switching back to single-pulse regime, which we here call defragmentation. Defragmentation also temporarily requires a supply of energy to the intra-cavity field, which is possible by self-organization of the internal dynamics of the intra-cavity field so as to minimize the absorption induced by the saturable absorber (SESAM) while increasing the intra-cavity gain (which depends not only on the PP but also on the average power of the field at the input of the EDF fiber). In this internal dynamic, the intra-cavity field can also adopt a spectral profile that minimizes the attenuation produced by the BPF. To gain insight into the role of the BPF in a defragmentation process, we have examined the evolution of the pulse parameters during the 2D process considered in the panels (a1) and (a2) of Fig. 8, in which the second step of the 2D process begins just after the fragmentation of the pulse generated in the first step of the process. Figure 10 shows the pulse parameters at the filter input as a function of the PP in the first phase of the process, and as a function of the filter bandwidth in the second phase of the process. The following general characteristics are observed: (i) Pulse parameters change gradually (helped in that because we vary the configuration parameters gradually to keep the evolution adiabatic), except at two specific moments which are nothing other than the times when fragmentation and defragmentation occur. As the fragmentation phenomenon has already been addressed as part of the conventional approach, hereafter we focus on the defragmentation process. Figure 10 shows that during the first phase of the process, some pulse parameters grow monotonically (Energy, peak power, and spectral width) while the temporal width decreases monotonically. At the end of the first phase, fragmentation occurs and causes an abrupt variation in all pulse parameters. It is quite remarkable that, in the second phase of the process, all pulse parameters continue to vary in the same direction as in the first phase, but in a more pronounced way. Thus, the pulse spectrum, which widened only moderately in the first phase of the process, continues to broaden in the second phase but at a faster rate, with the peculiarity that this rate is higher than the rate of widening of the BPF bandwidth (indicated by the dashed lines in panels 10 (d1) and10 (d2)). Consequently, in the cavity equipped with Gaussian filter, the pulse spectrum, which was initially within the bandwidth limits, widens to the bandwidth limits. Defragmentation occurs when the pulse spectrum begins to overflow the bandwidth, as indicated by the small arrow in panel 10 (d1). The action of the filter, which is then at its peak, completely destabilizes the multi-pulse state and causes a restructuring of the intensity profile resulting in a containment of the pulse spectrum within the bandwidth limits. In fact, this restructuring of the pulse profile corresponds to a relaxation process because the resulting spectral containment is so strong that the resulting pulse is in a state where the influence of the filter is negligible. The relaxation process also occurs in the cavity equipped with the flat-top filter, as can be seen in panel 10 (d2), but it is preceded by a phenomenology that differs drastically from that described for the Gaussian filter. Indeed, at the beginning of the second phase of the 2D procedure (i.e. at $\Delta\lambda$=8nm) in the cavity equipped with the flat-top filter, the pulse spectrum is already slightly overflowing from the filter bandwidth. As $\Delta\lambda$ increases, this overflow increases to the peak of the filter destabilizing

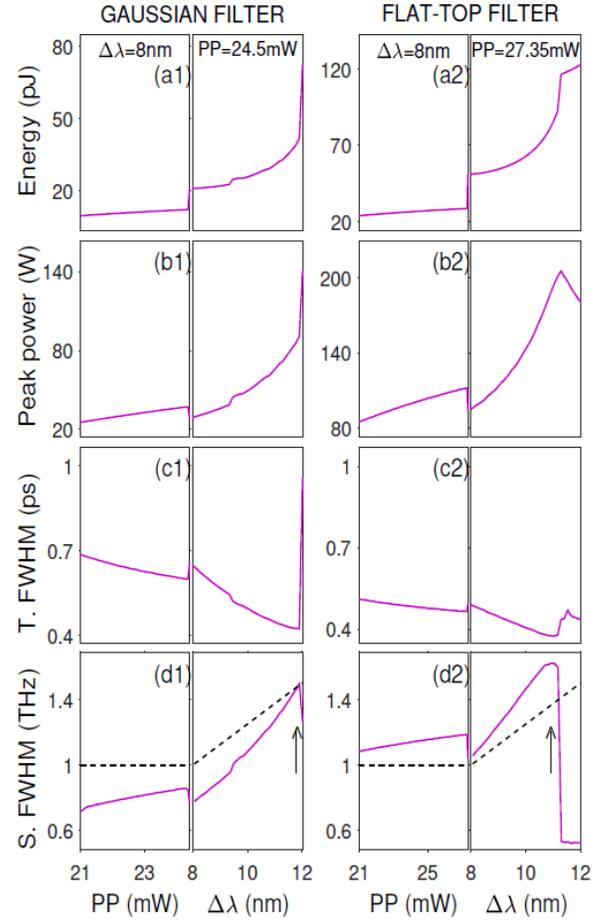

F                                           2D procedure shown in panels (a1) and (a2) of Fig. 8. (a1)-(a2) : Pulse energy. (b1)-(b2) : Pulse peak power. (c1)-(c2) : Temporal FWHM (d1)-(d2) : Spectral FWHM.

effect, where defragmentation occurs.

Furthermore, it is worth noting in Figs. 8 and 9 that the 2D approach under study has an obvious interest which is to significantly reduce the level of PP required to access the multi-pulse regime. Another major advantage of this approach is the multitude of routes it offers to access the multi-pulse regime. For example, Fig. 11 shows the result we get by following a different route from that of Fig. 8, to access the multi-pulse regime for $\Delta\lambda$ = 12nm. In Fig. 11 we use a 2D approach where the first step starts with $\Delta\lambda_{opt}$ = 4nm. Panels 11 (a1) and 11 (a2) show the result obtained when the first step of the procedure ends just after the fragmentation point, i.e., at a PP of 25.45mW (30mW) for the Gaussian (flat-top) filter. Panels 11 (b1) and 11 (b2) show the result obtained for a PP slightly above the fragmentation point, while panels 11 (c1) and 11 (c2) correspond to the case where the PP is slightly below the fragmentation point. Fig. 12 gives a better overview of the intensity profile of the multi-pulse structure generated at the end of the 2D procedure displayed in panels (b1) and (b2) of Fig. 11. It is clear that the results of figures 11 and 12 have the

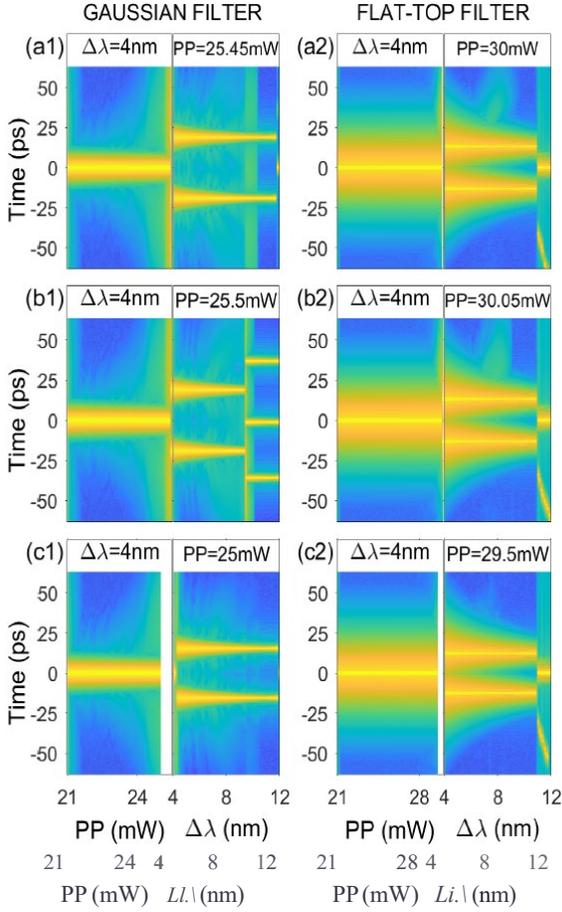

FIG. 11. Evolution of the 2D contour of the temporal intensity as a function of the pumping power for $\Delta\lambda = 4$ nm (first step), and according to the bandwidth of the filter (second step). (a1)-(a2) : The final PP is set just after the fragmentation point. (b1)-(b2) : The final PP is set slightly above the fragmentation point. (c1)-(c2) : The final PP is set slightly below the fragmentation point.

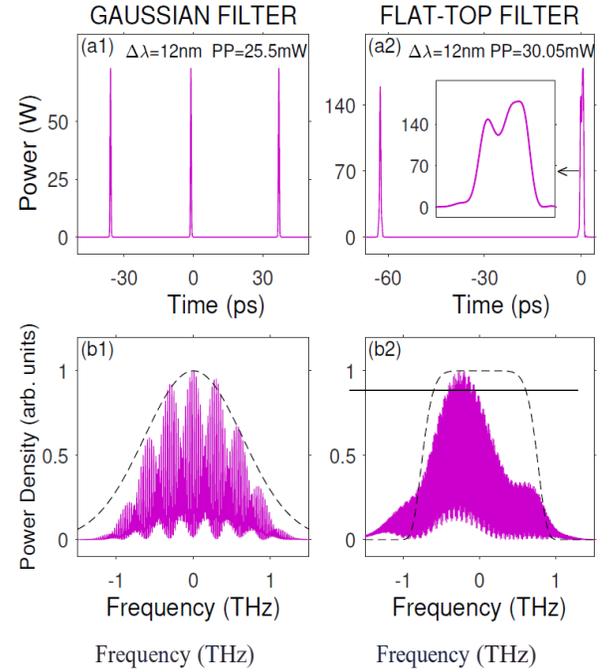

FIG. 12. Multi-pulse intensity profile generated at the end of the two-dimensional procedure shown in panels (b1) and (a2) of Fig. 11. (a1)-(a2) : Temporal profile. (b1)-(b2) : Spectral profile.

same general characteristics as those of Fig. 8 and 9, as mentioned below:

(i) Three (two) pulses are generated in the cavity equipped with the Gaussian (flat-top) filter.

(ii) The pulses generated with the Gaussian filter are physically identical while those generated with the flat-top filter have different characteristics.

(iii) The spectrum of the final multi-pulse structure occupies almost the entire bandwidth of the filter, as shown in panels 9(b1)-9(b2) and 12(b1)-12(b2), thus indicating that, regardless of its bandwidth and its spectral pro- file, the filter plays a decisive role in the 2D procedure of access to the multi-pulse regime whereas in the conventional approach the filter plays almost no role when $\Delta\lambda > \Delta\lambda_c$ Thus, the aforementioned general characteristics clearly indicate that the two routes we have chosen to access the multi-pulse regime lead to two states located in the immediate vicinity of the same fixed point (of the laser dynamics), with which they share the same general characteristics. The two states show only small differences in their temporal profiles, because of the slight difference in the respective PPs used in the two routes to the multi-pulse regime.

In the following, we will focus on a striking simulation result appearing in the figures 9 and 12, namely, the asymmetry between the two pulses generated by the cavity equipped with flat-top filter after the filter bandwidth has been increased to 12nm. Indeed, we initially expect that the shape of all pulses traveling round the laser cavity would be identical, defined by a common dissipative soliton attractor [9]. The simultaneous propagation of pulses with different profiles is rather exceptional, requiring specific laser cavity parameters [47]. Note that in the later case, pulses having different profiles are expected to travel at different velocities, unless they form a compact composite soliton molecule. A close examination of the second phase of the 2D process displayed in figures 8 and 11 leads us to attribute this asymmetry between the two pulses to the hybridization of the defragmentation process with other phenomena involved in laser dynamics. Indeed, we define a perfect defragmentation process as a process that generates a state where the pulses are less numerous, but all identical. On the other hand, we see in figures 8 and 11 that after this process of *hybrid defrag- mentation,* which occurs for a value of $\Delta\lambda$ between 11nm and 12nm, the less intense pulse drifts in the temporal domain monotonically as that $\Delta\lambda$ approaches 12nm. We attribute this temporal drift to the combined action of several phenomena, among which SRS (stimulated Ra- man scattering) plays the key role [46, 48, 49]. SRS comes into play because of the relatively high peak power of the two pulses (100W) and the difference in peak power between the two pulses (as can be seen in the panels (a2) of figures 9 and 12). One of the peculiarities of the 2D approach for accessing the multi-pulse regime is that the BPF plays a crucial role not only when its bandwidth $\Delta\lambda$ is less than $\Delta\lambda_c$, but also when $\Delta\lambda > \Delta\lambda_c$. Despite its lumped-filtering effect, a clear qualitative picture of the BPF contribution to the intra-cavity dynamics can be obtained using a collective-coordinate approach in which the lumped filtering is approximated by an equivalent continuous distributed filtering action along the cavity, defined by $F[\psi] \equiv \xi \psi_{tt}$, where $\xi$ is a constant that de- pends on the spectral profile of the BPF and the length of the cavity [48, 49]. On the other hand, by neglecting the TOD and using the linear approximation of $\tilde{\chi}_R$, the NLSE takes the following form [48, 49] :

$$\psi_z + \frac{i\,\beta_2(z)}{2}\,\psi_{tt} + \frac{\alpha(z)}{2}\psi - i\gamma\,|\psi|^2\,\psi$$
$$= \xi\psi_{tt} + i\,\gamma\,\rho\,f_1\,\psi\frac{\partial|\psi|^2}{\partial t}. \qquad (3)$$

The last term in Eq. (3) represents the Raman contribution, with $f_1 \approx 7e-3$ as a typical value [49]. To obtain a qualitative idea of the internal dynamic of the pulse, we assume a Gaussian ansatz given by :

$$\psi_g = x_1\,exp(-\eta^2/x_3^2 + ix_4\eta^2/2 + i\,x_5\eta + ix_6), \qquad (4)$$

where $\eta = t - x_2$, and where $x_1$, $x_2$, $2\sqrt{(ln2)}x_3$, $x_4/(2\pi)$, $x_5/(2\pi)$ and $x_6$ represent the pulse's amplitude, temporal position, width (FWHM), chirp, frequency, and phase, respectively. Then, following a collective-variable approach [49], we obtain the following explicit analytical expressions for the dynamics of the pulse parameters :

$$\dot{x_1} = -\alpha x_1/2 + \beta_2 x_1 x_4/2 - x_1(x_5^2 + 2x_3^{-2})\xi \qquad (5)$$

$$\dot{x_2} = -\beta_2 x_5 - x_4 x_3^2 x_5 \xi \qquad (6)$$

$$\dot{x_3} = -\beta_2 x_3 x_4 + (4 - x_4^4 x_2^2)x_3^{-2}\xi/2 \qquad (7)$$

$$\dot{x_4} = -(4 - x_4^4 x_4^2)x_3^{-4}\beta_2 - \sqrt{2}x_1^2 x_3^{-2}\gamma - 8x_4 x_3^{-2}\xi \quad (8)$$

$$\dot{x_5} = -\sqrt{2}\gamma_r x_1^2 x_3^{-2} - x_5(4x_3^{-2} + x_3^2 x_4^2)\xi \qquad (9)$$

$$\dot{x_6} = -\frac{\beta_2}{2}(x_5^2 - 2x_3^{-2}) + \frac{5\sqrt{2}\gamma}{8}x_1^2 + (1 - x_3^2 x_5^2)x_4\xi. \qquad (10)$$

where $\gamma_r = \rho f_1$. Let us consider the internal dynamics of a pulse, where the initial condition is chosen to be a free- chirp point of the cavity, i.e., $x_4(z=0) = 0$. Suppose that at $z=0$, the pulse's spectrum is located exactly at the center of the bandwidth of the system, i.e. with no frequency shift : $x_5(z=0) = 0$. Then, the first term on the r.h.s of Eq. (9) indicates that SRS causes the pulse spectrum to drift continuously towards low frequencies at a rate proportional to the pulse's peak power ($x^2$) and inversely proportional to the square of its temporal width ($x^2$). The second term on the r.h.s of Eq. (9), which is proportional to the resulting frequency shift $x_5 = 0$, triggers one of the filtering effects, which is to oppose any frequency drift. Thus, the spectral filter imposes a limit on the drift caused by SRS, but does not bring the pulse's spectrum back to the center of the bandwidth. When this limit is reached, the corresponding spectral shift is converted into a temporal shift by the chromatic dispersion of the intracavity fiber system, as shown by the first term on the right-hand side of Eq. (6). How- ever, in a dispersion-compensated cavity such as the one considered in the present study, the temporal shifts ($x_2$) occur in opposite directions, $x_2 > 0$ or $x_2 < 0$, depending on whether the pulse propagates in a fiber section with positive ($\beta_2 > 0$) or negative ($\beta_2 < 0$) dispersion; which should lead globally to a zero temporal shift after a complete cavity round-trip. The second term in the r.h.s of Eq. (6) shows that the filter also converts the pulse's frequency shift into a temporal shift, but the cavity has no mechanism to compensate for this temporal shift. Thus, the difference in peak power between the two pulses generated in the cavity equipped with the flat-top filter [see Figs. 9 (a2) and 12 (a2)] leads to a difference in their respective Raman-induced frequency shifts, which is con- verted into a difference in their respective drift speeds in the temporal domain

## V. CONCLUSION

In this study we have analyzed the strategies of access to the multi-pulse regime of a MLFL, and showed that the operational conditions for generating stable multi- pulse structures can be dramatically improved by including a tunable-bandwidth spectral filter within the laser cavity. Our analysis shows that an adequate spectral filter setting results in a drastic reduction in the PP needed to access the multi-pulse regime, of typically 50% for the laser cavity configuration we have considered. We have shown that the conventional procedure of reaching the multi-pulse operation through the sole increase of the PP is effective only when the filter bandwidth ($\Delta\lambda$) is small enough, i.e., below a critical value ($\Delta\lambda_c$) which depends on the configuration of the other cavity components. On the other hand, for $\Delta\lambda > \Delta\lambda_c$ the conventional method is inefficient because the level of PP required to generate multi-pulse structures is significantly higher than that of the region where $\Delta\lambda < \Delta\lambda_c$, and the intensity profiles of the generated pulses are affected by important distortions with respect to a bell-shaped profile.

One of the most important results of the present study is the development of a two-dimensional (2D) parameter adjustment procedure that avoids the disadvantages of the conventional approach aroused. To generate multi- pulse structures via the 2D approach with a bandwidth $\Delta\lambda > \Delta\lambda_c$, we proceed in two steps. Firstly, we set the bandwidth value to $\Delta\lambda_{opt} < \Delta\lambda_c$, and we gradually in- crease the PP to a level slightly above the fragmentation point. At the end of this step, the laser is in multi-pulse regime, with a relatively low PP, but with a bandwidth $\Delta\lambda_{opt}$ which is not the desired bandwidth. The second step of the procedure is to gradually increase the bandwidth up to the desired value $\Delta\lambda$, by keeping the PP at a level slightly above the fragmentation point. Using this two-dimensional method we have succeeded in generating multi-pulse structures with distortion-free pulse profiles, and with a PP level as low as those required in the region where $\Delta\lambda < \Delta\lambda_c$.

Furthermore, the present study reveals major qualitative differences in the intensity profile of the stable states, depending on whether the cavity is equipped with a Gaussian or flat-top filter. The flat-top filter promotes the generation of pulses with higher peak power, which in turn exacerbate non-linear effects and instability phenomena such as defragmentation, distortions of pulse in- tensity profile, and pulse temporal drift phenomena induced by Raman scattering and chromatic dispersion of intra-cavity fibers. Despite the increase in the complexity of pulse dynamics in the latter case, we have shown that the analysis of the dynamics of multi-pulse structures remains possible using collective coordinate methods.

Finally, our 2D approach of the present work opens up new possibilities for access to various types of multi- pulse structures, including bound states of solitons such as soliton molecules, macromolecules, complex molecular or soliton crystals.

*We acknowledge support from the Conseil r´egional de Bourgogne-Franche-Comt´e, iXCore Research Foundation and Agence Nationale de la Recherche (through projects ANR-15- IDEX-0003 and ANR-17-EURE-0002).*

---